\documentclass{article}

 \usepackage[eandd,final]{neurips_2026}

\usepackage[utf8]{inputenc} 
\usepackage[T1]{fontenc}    
\usepackage{hyperref}       
\usepackage{url}            
\usepackage{booktabs}       
\usepackage{amsfonts}       
\usepackage{nicefrac}       
\usepackage{microtype}      
\usepackage{multirow}
\usepackage{multicol}
\usepackage{booktabs}
\usepackage{pifont}   
\usepackage{color}    
\usepackage{graphicx}
\graphicspath{{./}{../}}
\usepackage{subcaption}
\usepackage[table]{xcolor}
\usepackage[most]{tcolorbox}
\tcbuselibrary{listings,breakable}
\definecolor{promptcolor}{RGB}{255,245,245}
\newtcblisting{promptbox}[1][]{
  enhanced,
  listing only,
  colback=promptcolor,
  colframe=red!70!black,
  boxrule=0.8pt,
  arc=3pt,
  left=3pt, right=3pt, top=2pt, bottom=2pt,
  listing options={
    basicstyle=\ttfamily\scriptsize,
    breaklines=true,
    columns=fullflexible,
    keepspaces=true,
    showstringspaces=false,
    aboveskip=0pt,
    belowskip=0pt
  },
  title=#1
}
\newcommand{\cmark}{\textcolor[rgb]{0, 0.6, 0}{\ding{51}}} 
\newcommand{\xmark}{\textcolor[rgb]{0.8, 0, 0}{\ding{55}}} 
\title{RepoZero: Can LLMs Generate a Code Repository from Scratch?}

\author{%
  Zhaoxi Zhang\thanks{Work done during internship at Baidu Inc.} \\
  Peking University\\
  \texttt{zhaoxizhang25@stu.pku.edu.cn} \\
  \And
  Yiming Xu \\
  Peking University \\
  \texttt{teddyxu@stu.pku.edu.cn} \\
  \AND
  Jiahui Liang \\
  Baidu Inc \\
  \texttt{liangjiahui03@baidu.com} \\
  \And
  Weikang Li \thanks{Correspondence Author}\\
  Baidu Inc \\
  \texttt{wavejkd@pku.edu.cn} \\
  \And
  Xiaoshuai Chen\\
  Independent Researcher \\
  \And
  Liwei Qian \\
  Baidu Inc \\
  \texttt{qianliwei@baidu.com} \\
  \And
  Xin Pei \\
  Baidu Inc \\
  \texttt{peixin@baidu.com} \\
  \And
  Jizhou Huang \\
  Baidu Inc \\
  \texttt{huangjizhou01@baidu.com} \\
  \And
  Rui Sun \\
  Independent Researcher \\
  \texttt{sunr138@pku.org.cn} \\
  \And
  Yunfang Wu \footnotemark[2]\\
  Peking University \\
  \texttt{wuyf@pku.edu.cn} \\
}

\begin{document}

\maketitle

\begin{abstract}
Large Language Models (LLMs) have recently shown remarkable progress in code generation, yet their ability to construct complete software repositories from scratch remains poorly understood. A fundamental bottleneck is the lack of verifiable and scalable evaluation: existing benchmarks either focus on patch-based editing or rely on human or LLM-based judgments, which introduce bias and limit reproducibility.
In this work, we present RepoZero, the first benchmark that enables fully automated, execution-based verification of repository-level generation from scratch. Our key idea is to reformulate generation as repository reproduction: given only API specifications, an agent must re-implement an entire repository such that its behavior matches the original implementation. This design allows for strict black-box validation via output equivalence, while naturally supporting large-scale construction by reusing existing open-source repositories. To further mitigate data leakage and shortcut solutions, we introduce cross-language constraints and a sandboxed evaluation protocol.
Building on this benchmark, we propose an Agentic Code-Test Evolution (ACE) framework that performs iterative test generation and error-driven refinement, enabling effective test-time scaling for repository-level synthesis.
Extensive experiments across multiple state-of-the-art LLMs and agent frameworks reveal that even the strongest LLM agents achieve only limited pass rates (30\% - 55\%), exposing a substantial gap between current capabilities and real-world software development requirements.
Our results establish RepoZero as a challenging, scalable, and reliable testbed for end-to-end code generation, and highlight self-verification via test generation as a critical direction for advancing LLM-based coding agents. Code is available at \href{https://www.example.com}{https://github.com/JesseZZZZZ/RepoZero}.
\end{abstract}

    
\begin{table*}[htbp]
\small
  \centering
  \caption{Comparison with existing coding benchmarks on multiple dimensions. \textbf{Repository} (generating a repository rather than a single code snippet), \textbf{Multi-lang.} (multiple programming languages), \textbf{From-scratch Gen.} (generation with no pre-given code), \textbf{Execution} (verifying the executed results), \textbf{Test Suites} (test cases rather than LLM-as-judge), \textbf{Scalability} (automatic pipeline rather than human-required), and \textbf{Leakage Defense} (preventing data leakage).}
  \label{tab:benchmark-comparison}
  \resizebox{0.99\textwidth}{!}{
  \begin{tabular}{lccccccc}
    \toprule
    \textbf{Benchmark} & \textbf{Repository} & \textbf{Multi-lang.} & \textbf{From-scratch Gen.} & \textbf{Execution} & \textbf{Test Suites} & \textbf{Scalability} &\textbf{Leakage Defense}\\
    \midrule
    HumanEval \citep{chen2021evaluating}& \xmark & \xmark & \xmark & \cmark & \cmark & \xmark &\cmark \\
    SWE-bench \citep{jimenez2024swebenchlanguagemodelsresolve}       & \cmark & \xmark & \xmark & \xmark & \cmark & \xmark  &\xmark \\
    Multi-SWE-bench \citep{zan2025multiswebenchmultilingualbenchmarkissue} & \cmark & \cmark & \xmark & \xmark & \cmark & \xmark &\xmark \\
    FEA-bench \citep{li-etal-2025-fea}       & \cmark & \xmark & \xmark & \xmark & \cmark & \xmark &\xmark \\
    EvoCodeBench \citep{zhang2026evocodebenchhumanperformancebenchmarkselfevolving}    & \cmark & \xmark & \cmark & \xmark & \cmark & \xmark &\xmark \\
    CodeS \citep{zan2024codes}           & \cmark & \xmark & \cmark & \xmark & \xmark & \xmark &\xmark \\
    E2EDevBench \citep{liu2025e2edev}     & \cmark & \xmark & \cmark & \xmark & \xmark & \xmark &\cmark \\
    NL2Repo \citep{ding2025nl2repo}     & \cmark & \xmark & \cmark & \cmark & \cmark & \xmark &\xmark \\
    \midrule
    \rowcolor{gray!20}
    \textbf{RepoZero (Ours)}   & \cmark & \cmark & \cmark & \cmark & \cmark & \cmark &\cmark \\
    \bottomrule
  \end{tabular}
  }
\end{table*}
\section{Introduction}
\label{sec:intro}
Recent advances in LLM-based coding have drawn increasing attention toward fully autonomous end-to-end software development. State-of-the-art large language models \citep{liu2025deepseek,yang2025qwen3,team2026kimi,xiao2026mimo,zeng2026glm} have been integrated into coding agents \citep{wang2024openhands,gao2025trae,zhang2025one} capable of completing complex programming tasks with minimal or no human intervention. From the perspective of benchmark difficulty, coding tasks can generally be categorized along two orthogonal dimensions. The first dimension concerns the scope of code modification, including: (1) single-file coding and (2) repository-level coding. The second dimension concerns the nature of the programming task itself, including: (1) debugging, (2) incremental development, and (3) generation from scratch.

Current coding agents have largely mastered single-file editing. However, evaluating repository-level modifications remains a burgeoning frontier. Benchmarks such as SWE-bench \citep{jimenez2024swebenchlanguagemodelsresolve}, Multi-SWE-bench \citep{zan2025multiswebenchmultilingualbenchmarkissue}, and FEA-bench \citep{li-etal-2025-fea} have pioneered the assessment of repo-level edits and incremental development in real-world scenarios. While these benchmarks leverage patch-based verification and unit testing, the domain of from-scratch repository generation remains relatively unexplored. A primary obstacle is the scarcity of pre-existing unit tests available online for newly generated projects. Consequently, current benchmarks for from-scratch generation, such as CodeS \citep{zan2024codes}, EvoCodeBench \citep{zhang2026evocodebenchhumanperformancebenchmarkselfevolving}, and NL2RepoBench \citep{ding2026nl2repobenchlonghorizonrepositorygeneration}, predominantly rely on human-in-the-loop or LLM-as-a-judge frameworks. These methods, however, introduce subjective biases from both human evaluators and language models, leading to potentially unreliable evaluation outcomes.

Data leakage presents another formidable challenge for repository-level coding benchmarks. Since modern large language models (LLMs) are extensively pre-trained on massive GitHub corpora \citep{zeng2026glm,xiao2026mimo}, it is increasingly difficult to discern whether their performance stems from genuine reasoning capabilities or mere rote memorization. While some studies \citep{zhang2026evocodebenchhumanperformancebenchmarkselfevolving} attempt to mitigate this by utilizing rare or recently published repositories, this approach inherently constrains the dataset scale, thereby precluding large-scale evaluation. Furthermore, such "clean" datasets are subject to rapid contamination post-publication, necessitating frequent and labor-intensive updates to maintain their integrity.

To address these challenges, we propose \textbf{RepoZero}, the first verifiable benchmark specifically designed to evaluate the from-scratch repository-level generation capabilities of LLM agents. RepoZero achieves rigorous verification by reformulating repository-level generation as a repository reproduction task. Specifically, for a target repository (e.g., a pip-installed Python package), we first employ an LLM to generate test files that invoke the repository's APIs; for each test file, the LLM generates several test cases, which are then filtered to retain only successful test cases as ground truth. An agent is then tasked with generating an entire target repository that replicates the source repository's functionality and API signatures. Evaluation is performed by executing the generated target repository against the saved test cases and comparing the outputs with those of the source repository. To mitigate data leakage, we mandate cross-language synthesis--requiring agents to implement the repository in a different programming language--and prohibit the use of external site packages. Furthermore, we introduce an Agentic Code-Test Evolution (ACE) workflow for test-time scaling. Unlike existing frameworks, where the lack of a verifiable environment makes retrieving ground-truth outputs for LLM-generated tests impossible, RepoZero leverages the source repository as an oracle, providing a deterministic and automated gold standard for evaluation.

In summary, our primary contributions are as follows:

1. \textbf{Introduction of RepoZero}: We establish the first verifiable and highly scalable benchmark tailored for repository generation from scratch, bridging the evaluation gap in existing repo-level coding tasks.

2. \textbf{Test-Time Scaling Framework}: We propose a novel agentic framework centered on test-time scaling, which significantly enhances the success rate of complex, repository-level synthesis.

3. \textbf{Comprehensive Benchmarking}: We conduct an extensive evaluation of state-of-the-art models and agentic scaffolds, providing critical insights and a robust foundation for future research in autonomous software engineering.

    
\section{Benchmark: RepoZero}
\subsection{Statistics of RepoZero}
\begin{figure}[t]
    \centering
    \includegraphics[width=1\linewidth]{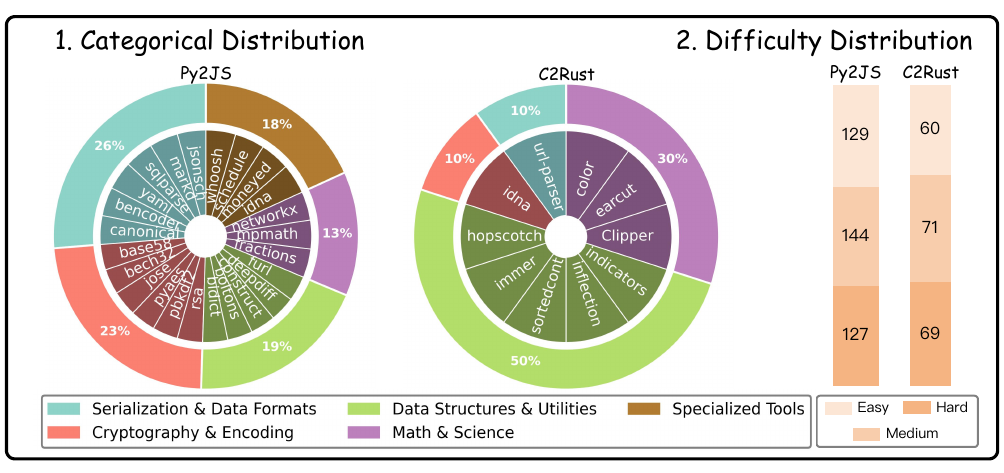}
    \caption{Demonstration of (1) categorical distribution, and (2) difficulty distribution of RepoZero-Py2JS (left) and RepoZero-C2Rust (right).}
    \label{fig:distribution}
\end{figure}

RepoZero consists of two subsets: \textbf{RepoZero-Py2JS}, where the source repositories are implemented in Python, and the agent is required to reimplement them in JavaScript, and \textbf{RepoZero-C2Rust}, where the source repositories are written in C/C++ and the agent is tasked with reimplementing them in Rust. As illustrated in Fig.~\ref{fig:distribution}, RepoZero covers a diverse range of repository categories. All repositories are manually curated and further preprocessed following the workflow described in Sec.~\ref{sec:construction}.
We employ LLMs to estimate the difficulty of each sample under several rubrics: Lines of Code (LOC), API counts, and input\slash output complexity. We apply majority voting (Claude-4.6-Opus, GLM-5, and DeepSeek-V3.2) to obtain the final estimation. The datasets are accordingly partitioned into \textit{Easy}, \textit{Medium}, and \textit{Hard} subsets. From the curated repositories, we construct 400 samples for RepoZero-Py2JS and 200 samples for RepoZero-C2Rust. The source repositories of these samples are manually selected. Because the remaining pipeline is automated after repository selection, the benchmark can be continuously updated to mitigate potential data leakage (once released, RepoZero will inevitably be collected as part of the training data of foundational models).
\subsection{Task Definition and Challenges}
In a repository reproduction task, an LLM agent is provided with: (1) the functional specifications of the source repository's APIs, (2) four white-box test cases (comprising inputs, ground-truth outputs, and descriptions), and (3) constraints to preclude "shortcut" behaviors, such as cross-language embedding or direct API delegation. The target repository is compiled, where applicable, and evaluated via black-box testing to ensure its outputs strictly align with the source repository's baseline.
This task presents four primary challenges for LLM agents: (1) Functional Synthesis, translating high-level API specifications into concrete implementations; (2) Modular Reasoning, resolving inter-module dependencies across multiple files; (3) Iterative Self-Correction, refining code through autonomous test design and execution; and (4) Long-context Retention, maintaining architectural coherence over extensive reasoning horizons. As shown in Sec.~\ref{sec:results}, current agents exhibit several limitations in navigating these complexities.
\subsection{Evaluation Pipeline}
We employ two popular coding agents: OpenHands-bash \citep{wang2024openhands,sutawika2026codescout} and Mini-SWE-Agent \citep{yang2024sweagent}. These agents can interact with the computer system via a series of tools, for instance, a terminal, where the agent can run commands. The agent is prompted with the task information and prompted to generate a repository with an assigned entry point. After finishing the generation, black-box tests are applied to test the quality of the target repository.

To ensure a rigorous evaluation, we address the potential for "cheating" behaviors—such as accessing ground-truth repositories or manipulating test suites—frequently observed in recent LLM agent benchmarks \citep{jimenez2024swebenchlanguagemodelsresolve, merrill2026terminal}. We implement a multi-layered protocol to maintain the integrity of the evaluation process:

\paragraph{Black-box Testing}
To mitigate the risk of agents overfitting to specific test outcomes, we implement a black-box evaluation framework. Only four representative white-box test cases are included in the prompt for contextual guidance, while the remainder of the evaluation suite is withheld. By prioritizing semantic correctness over superficial execution results, this methodology ensures that agents cannot bypass rigorous verification through trivial output matching or heuristic shortcuts.

\paragraph{Environment Isolation and Security}
Evaluation is performed within isolated Docker containers characterized by restricted file-system permissions. To maintain the integrity of the benchmark, the source repository’s test files are configured as read-only, and the hidden evaluation suite is never exposed to the agent’s accessible file system. These safeguards prevent agents from manipulating or circumventing challenging test cases to artificially inflate performance metrics.

\paragraph{Execution Constraints and Cross-Language Integrity}
We impose stringent limitations on system-level commands to preclude non-generalizable solutions. Specifically, agents are prohibited from importing external packages into the target repository. Furthermore, we enforce strict language consistency: for instance, if an agent is tasked with porting a Python library (e.g., mpmath) to JavaScript, the use of bridge commands or cross-language invocations (e.g., executing Python scripts via shell wrappers) is strictly forbidden. Such constraints necessitate that agents rely on complex logical reasoning and cross-language synthesis rather than delegating tasks to external APIs.

\begin{figure*}[t]
    \centering
    \includegraphics[width=0.95\linewidth]{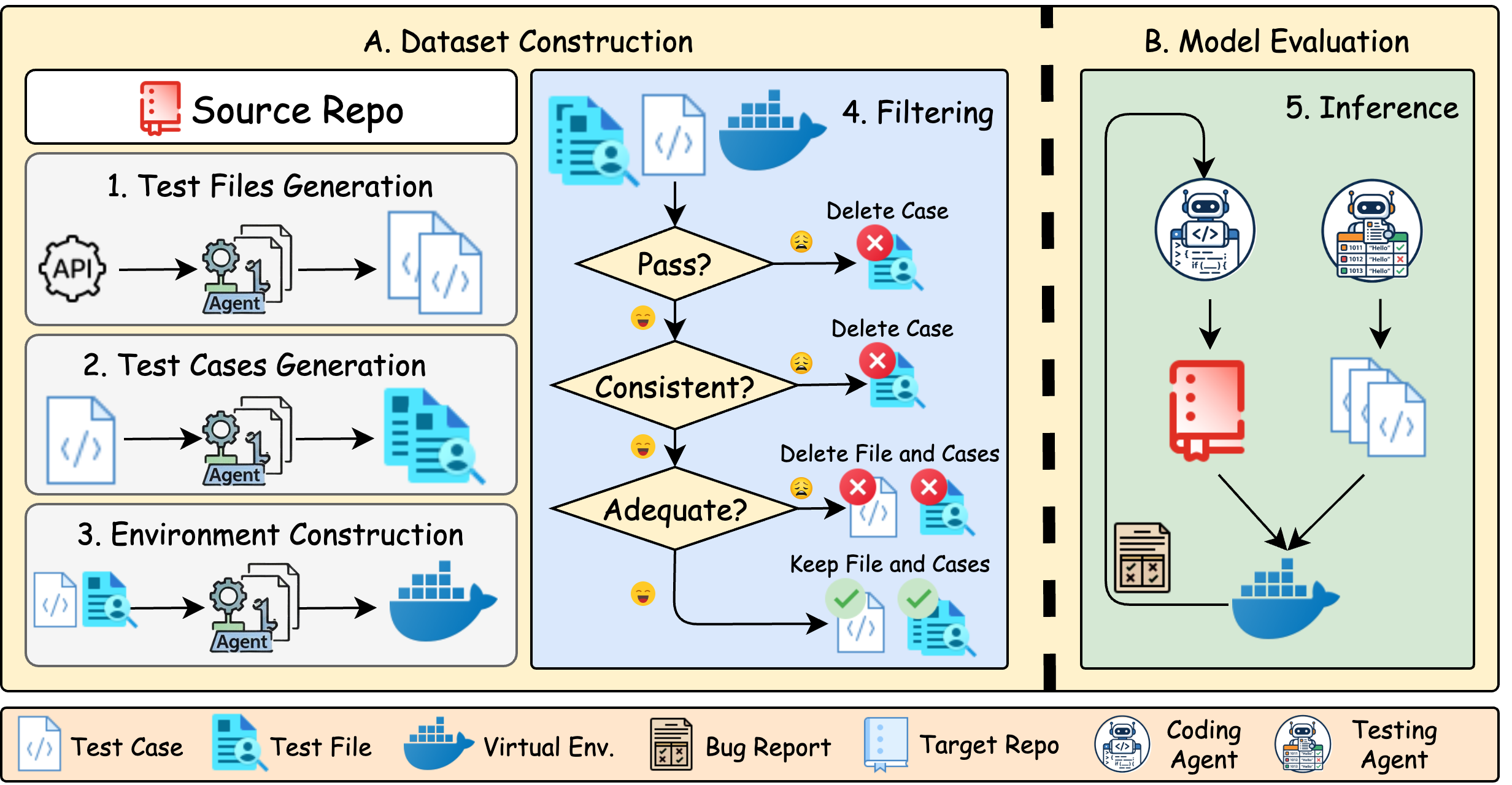}
    \caption{Illustration of (A) the construction of the RepoZero benchmark, and (B) agentic behavior during the evaluation stage. The loop ends when all  cases produced by the testing agent succeed.}
    \label{fig:main_figure}
\end{figure*}

\section{Benchmark Construction}
\label{sec:construction}
In this section, we illustrate the dataset construction process of our benchmark. First, we manually select several open-source repositories from GitHub, and we prompt an LLM agent to generate test files (each successful file stands for one sample in the dataset) that invoke APIs from the repository. Second, we prompt a test-case generator to generate test cases for each file. Third, we conduct filtering techniques for all test cases and test files.
\subsection{Sample Generation}
We manually curate a set of source repositories, denoted as $R_o$, and leverage an LLM to generate corresponding test files. To modulate task difficulty, we employ strict prompting to regulate the number of API calls per file, ranging from 1 to 20. We emphasize that repository selection is governed by human expert judgment based on three criteria.
\textbf{Determinism}: The repository must be deterministic to ensure consistent string-level comparisons.
\textbf{Open-source Integrity}: The repository must be fully open-source, excluding non-public components such as proprietary built-in hash tables or frozen model weights.
\textbf{Architectural Complexity}: The repository must be sufficiently complex to preclude single-file implementations, thereby rigorously evaluating repository-level coding capabilities.

\subsection{Test Case Generation}
\label{sec:test_case_gen}
We employ an LLM as a generator to produce evaluation datasets. For each target repository, the LLM is provided with the repository name and the corresponding test files to extract the underlying API specifications. Based on these contexts, the LLM is prompted to generate a comprehensive suite of input-output pairs. However, since the LLM has limited visibility into the entire codebase and the source repositories themselves may contain inconsistencies, the initial outputs are often unpredictable. Consequently, a rigorous filtering pipeline is essential to ensure the quality and reliability of the generated test cases.

\subsection{Environment Configuration}
To facilitate execution, we instantiate an isolated Docker container for each sample, equipped with standard development tools (e.g., Python, g++, etc.). We leverage OpenHands \citep{wang2024openhands} to dynamically resolve and install the dependencies required by the source test files. An environment is validated as "successful" only if it can execute the generated test files and pass at least five baseline cases without runtime errors. This stage ensures that the environment is correctly configured to support the subsequent large-scale verification of the entire test suite.

\subsection{Data Filtering and Ground-Truth Verification}
With the repositories, test files, and candidate test cases assembled, we perform a multi-stage filtering process to curate valid evaluation samples. Each test case is executed 20 times within its designated environment to ensure stability. Test cases are discarded if they trigger: (1) runtime exceptions, or (2) non-deterministic outputs. For example, outputs involving memory addresses (pointers) or time-dependent variables are eliminated due to their inherent volatility. This rigorous pruning ensures that each remaining test case is both executable and deterministic, with its stable output serving as the definitive ground-truth for model evaluation. Samples are discarded if the number of valid test cases is fewer than 10.

\section{Agentic Code-Test Evolution Workflow}

Recent studies have widely acknowledged that jointly generating test cases alongside code can substantially improve code generation performance \citep{ding2025nl2repo}. However, the code-test feedback loop has not been broadly adopted in existing workflows for two primary reasons. First, LLMs do not consistently exhibit the tendency to generate test cases without explicit prompting or a predefined procedural framework. Second, test cases generated by LLMs are typically not independently verifiable; consequently, such cases can only assess code executability rather than functional correctness.

Inspired by Agentless \citep{xia2024agentless}, which replaces fully autonomous agentic procedures with a structured workflow to enhance LLM-based bug localization, we propose the \textbf{Agentic Code-Test Evolution} (ACE) workflow, a dedicated framework designed for RepoZero. As illustrated in Fig.~\ref{fig:main_figure} (B), ACE adopts an iterative code--test feedback loop.

RepoZero provides a suitable testbed for evaluating ACE. This is primarily because the ground-truth outputs for RepoZero samples are obtained by executing the source repository itself, which enables the generation of an effectively unlimited number of test cases while ensuring the correctness and reliability of their corresponding labels.

In practice, ACE serves as a general framework that can be integrated with arbitrary coding agents. The workflow begins with a coding agent that generates a target repository from scratch based on the APIs of the source repository. Subsequently, a testing agent produces multiple test cases and executes the source repository to obtain the corresponding ground-truth outputs. Notably, the filtering strategy described in Sec.~\ref{sec:test_case_gen} is also applied at this stage. If the generated repository successfully passes all test cases, the iterative loop terminates. Otherwise, the resulting error messages are fed back to the coding agent, which then revises the generated repository accordingly.
\section{Experimental Setup}
\label{sec:setup}
\paragraph{Models and Scaffolds}
We evaluate our benchmark using two widely recognized agent scaffolds for software engineering: OpenHands-bash \citep{sutawika2026codescout} and Mini-SWE-Agent \citep{yang2024sweagent}. These frameworks are integrated with state-of-the-art LLMs, including Kimi-K2.5, Kimi-K2.6 \citep{kimi2026kimik26}, GLM-5, GLM-5.1 \citep{zeng2026glm}, DeepSeek-V3.1, DeepSeek-V3.2 \citep{deepseekv31}, DeepSeek-V4 \citep{deepseek2026deepseekv4}, Ernie-5.0 \citep{wang2026ernie}, Minimax-M2.5, Minimax-M2.7 \citep{minimax2026minimaxm27} and Claude-4.6 \citep{anthropic2026claudesonnet46}. To ensure maximum flexibility, we do not impose an explicit cap on the number of tool-calling iterations; the only constraint is the maximum output token limit inherent to the base models. Furthermore, agents are permitted full access to context management techniques and retrieval-augmented tools to facilitate complex problem-solving.

\paragraph{Metrics} 
We adopt the Pass Rate (PR) as our primary evaluation metric. A sample is considered successful ($PR=1$) if and only if the outputs from both the source and target repositories exhibit strict string-level consistency; otherwise, it is assigned a value of $0$. Additionally, we report the Success Rate (SR) across test cases. While the SR (see Appendix~\ref{sec:more_results}) offers a complementary perspective, the PR is utilized as the more robust and definitive measure of functional equivalence. Results are reported as mean $\pm$ bootstrap standard deviation.


\begin{table*}[t]
\centering
\caption{Evaluation with \includegraphics[height=0.9em]{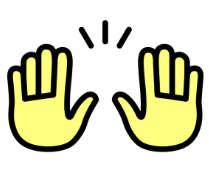} Openhands-bash on RepoZero. We present the pass rate of models across Easy, Medium, and Hard task difficulties.}
\label{tab:openhands}
\resizebox{0.89\textwidth}{!}{
\begin{tabular}{l|cccc|cccc}
\toprule
\multirow{2}{*}{\textbf{Model}} & \multicolumn{4}{c|}{\includegraphics[height=0.9em]{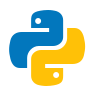} \textbf{Py2JS} \includegraphics[height=1em]{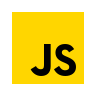}} & \multicolumn{4}{c}{\includegraphics[height=1em]{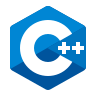} \textbf{C2Rust} \includegraphics[height=0.9em]{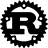}} \\ \cmidrule(lr){2-5} \cmidrule(lr){6-9} 
 & Easy & Medium & Hard & \textbf{Avg.} & Easy & Medium & Hard & \textbf{Avg.} \\ \midrule
\includegraphics[height=0.9em]{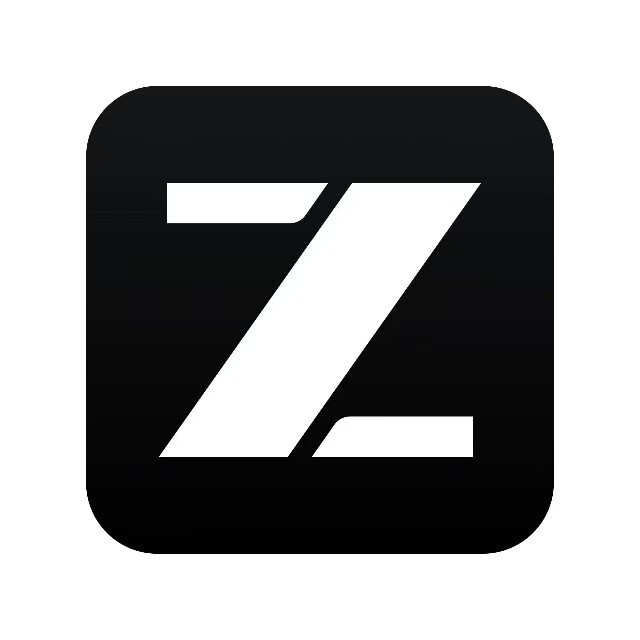} GLM-5 & 37.21 & 23.61 & 22.05 & 27.46$\pm2.29$ & 31.67& 20.00& 20.29& 23.47$\pm$2.34 \\
\includegraphics[height=0.9em]{figs/zhipu.jpg} GLM-5.1 & 46.51 & 38.89 & 16.54 & 34.19$\pm$2.38  & 45.90 & 31.43 & 28.99 & 35.05$\pm$2.19 \\
\includegraphics[height=0.7em]{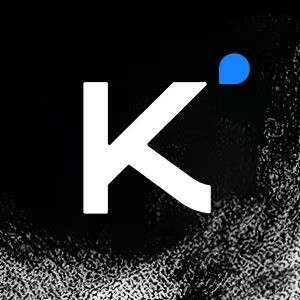} Kimi-K2.5 & 36.36 & 33.02 & 24.74 & 31.30$\pm$2.97  & 40.98& 30.00& 28.99& 33.00$\pm$1.97 \\
\includegraphics[height=0.7em]{figs/kimi.jpg} Kimi-K2.6 & 47.29 & 27.78 & 25.20 & 33.25$\pm$2.36  & 50.82& 30.00& 36.23& 38.51$\pm$2.03 \\
\includegraphics[height=0.9em]{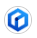} Ernie-5.0 &  27.91&  18.75&  11.81&  19.46$\pm$2.02 & 22.95 & 12.86 & 4.35 & 13.97$\pm$2.36  \\
\includegraphics[height=0.9em]{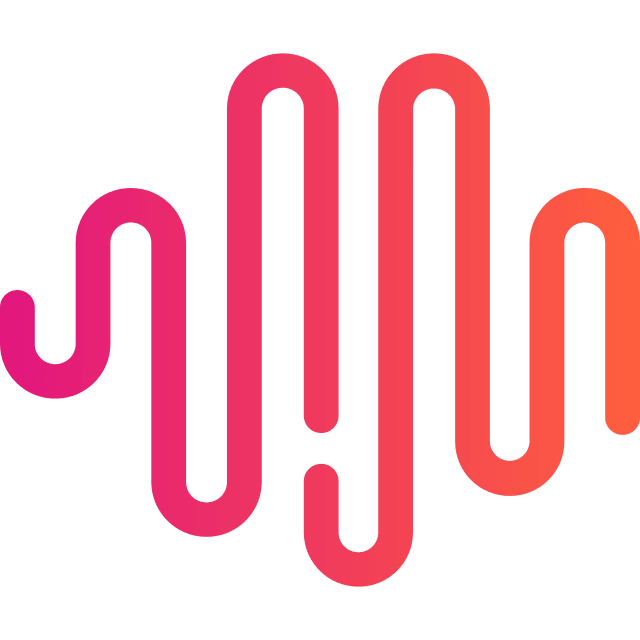} Minimax-M2.5 & 29.97 & 22.23 & 18.89 & 22.72$\pm$1.92  & 36.07 & 18.57 & 31.88 & 28.48$\pm$2.31  \\
\includegraphics[height=0.9em]{figs/minimax-color.png} Minimax-M2.7 & 34.88 & 29.17 & 16.54 & 27.15$\pm$2.15  & 37.70 & 25.71 & 33.33 & 32.12$\pm$2.91  \\
\includegraphics[height=0.9em]{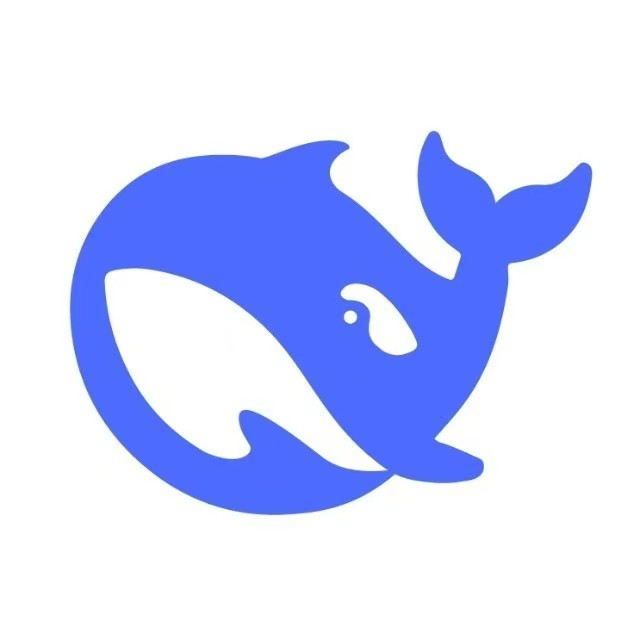} DeepSeek V3.1 & 34.11 & 25.69 & 18.11 & 26.08$\pm$2.26  & 43.33& 30.00& 26.09& 32.49$\pm$2.00 \\
\includegraphics[height=0.9em]{figs/deepseek.jpg} DeepSeek V3.2 & 37.21 & 31.25 & 18.11 & 29.01$\pm$2.34  & 48.33& 28.57& 26.09& 33.55$\pm$2.52 \\
\includegraphics[height=0.9em]{figs/deepseek.jpg} DeepSeek V4 Flash & 38.76 & 32.64 & 20.47 & 30.74$\pm$2.33  & 40.98 & 34.92 & 31.88 & 35.41$\pm$2.38  \\
\includegraphics[height=0.9em]{figs/deepseek.jpg} DeepSeek V4 Pro & 48.84 & 38.19 & 32.28 & 39.78$\pm$2.51  & 40.98 & 32.86 & 33.33 & 35.53$\pm$2.40  \\
\includegraphics[height=0.9em]{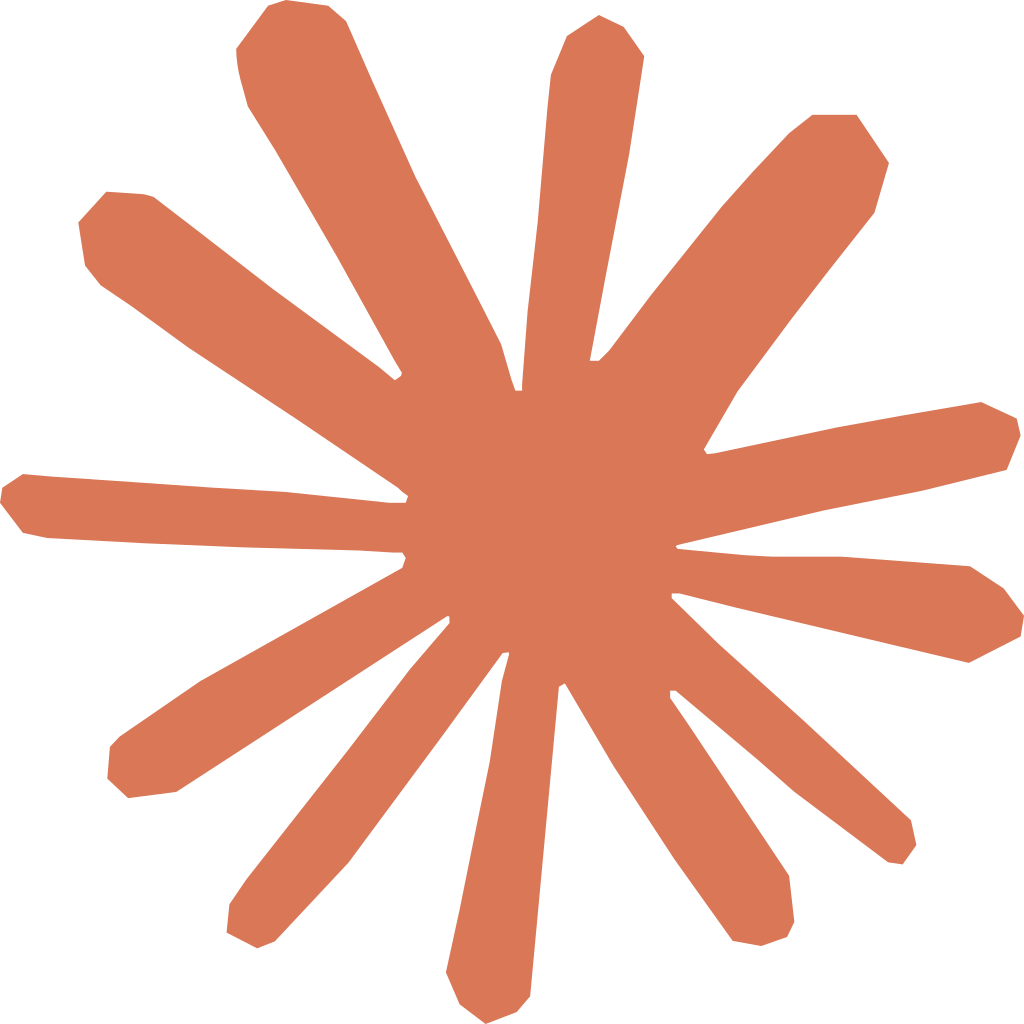} Claude-4.6-Sonnet &  59.69&  49.31&  45.67&  51.46$\pm$2.53 &  53.33&  45.71&  36.23&  44.60$\pm$2.74 \\

\bottomrule
\end{tabular}
}
\end{table*}
\begin{table*}[t]
\centering
\caption{Evaluation with \includegraphics[height=0.9em]{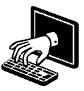} Mini-SWE-Agent on RepoZero. We present the pass rate of models across Easy, Medium, and Hard task difficulties.}
\label{tab:mini-swe-agent}
\resizebox{0.89\textwidth}{!}{
\begin{tabular}{l|cccc|cccc}
\toprule
\multirow{2}{*}{\textbf{Model}} & \multicolumn{4}{c|}{\includegraphics[height=0.9em]{figs/icons8-python-96.png} \textbf{Py2JS} \includegraphics[height=1em]{figs/icons8-javascript-96.png}} & \multicolumn{4}{c}{\includegraphics[height=1em]{figs/icons8-c-96.png} \textbf{C2Rust} \includegraphics[height=0.9em]{figs/icons8-rust-48.png}} \\ \cmidrule(lr){2-5} \cmidrule(lr){6-9} 
 & Easy & Medium & Hard & \textbf{Avg.} & Easy & Medium & Hard & \textbf{Avg.} \\ \midrule
\includegraphics[height=0.9em]{figs/zhipu.jpg} GLM-5 & 55.81 & 43.75 & 37.01 & 45.52$\pm$2.48  & 49.18& 37.14& 30.43& 38.51$\pm$1.98 \\
\includegraphics[height=0.7em]{figs/kimi.jpg} Kimi-K2.5 & 44.19 & 32.64 & 24.41 & 33.76$\pm$2.48  & 54.10& 40.00& 37.68& 43.53$\pm$2.33 \\
\includegraphics[height=0.9em]{figs/ernie.png} Ernie-5.0 & 34.11 & 15.28 & 13.39 & 20.77$\pm$2.08  & 26.23 & 14.29 & 17.39 & 19.97$\pm$2.04  \\
\includegraphics[height=0.9em]{figs/minimax-color.png} Minimax-M2.5 & 49.91 & 29.25 & 22.68 & 30.63$\pm$2.72  &  32.79&  24.29&  20.29&  25.39$\pm$2.31 \\
\includegraphics[height=0.9em]{figs/deepseek.jpg} DeepSeek V3.1 & 58.91 & 45.14 & 26.77 & 43.76$\pm$2.50  & 50.82& 37.14& 31.88& 39.45$\pm$2.51 \\
\includegraphics[height=0.9em]{figs/deepseek.jpg} DeepSeek V3.2 & 52.71 & 44.44 & 35.43 & 44.20$\pm$2.59  & 49.18& 38.57& 26.09& 37.41$\pm$2.78 \\
\includegraphics[height=0.9em]{figs/claude-color.png} Claude-4.6-Sonnet & 65.12 & 54.17 & 44.88 & 54.70$\pm$2.55  & 59.02 & 45.71 & 39.73 & 47.58$\pm$2.90  \\ \bottomrule
\end{tabular}
}
\end{table*}
\section{Results and Analysis}
\label{sec:results}
\subsection{Main Evaluation}
The primary experimental results for the RepoZero-Py2JS and RepoZero-C2Rust benchmarks are presented in Table~\ref{tab:openhands} and Table~\ref{tab:mini-swe-agent}, while results in Table~\ref{tab:distribution} illustrate the performance across multiple categories. Overall, all evaluated models exhibit suboptimal performance on these benchmarks regardless of whether the OpenHands-bash or Mini-SWE-Agent scaffold is employed. Notably, Claude-4.6-Sonnet consistently achieves the highest performance across all evaluation metrics.

\begin{figure}[b]
    \centering
    \begin{subfigure}[t]{0.44\textwidth}
        \centering
        \includegraphics[width=\linewidth]{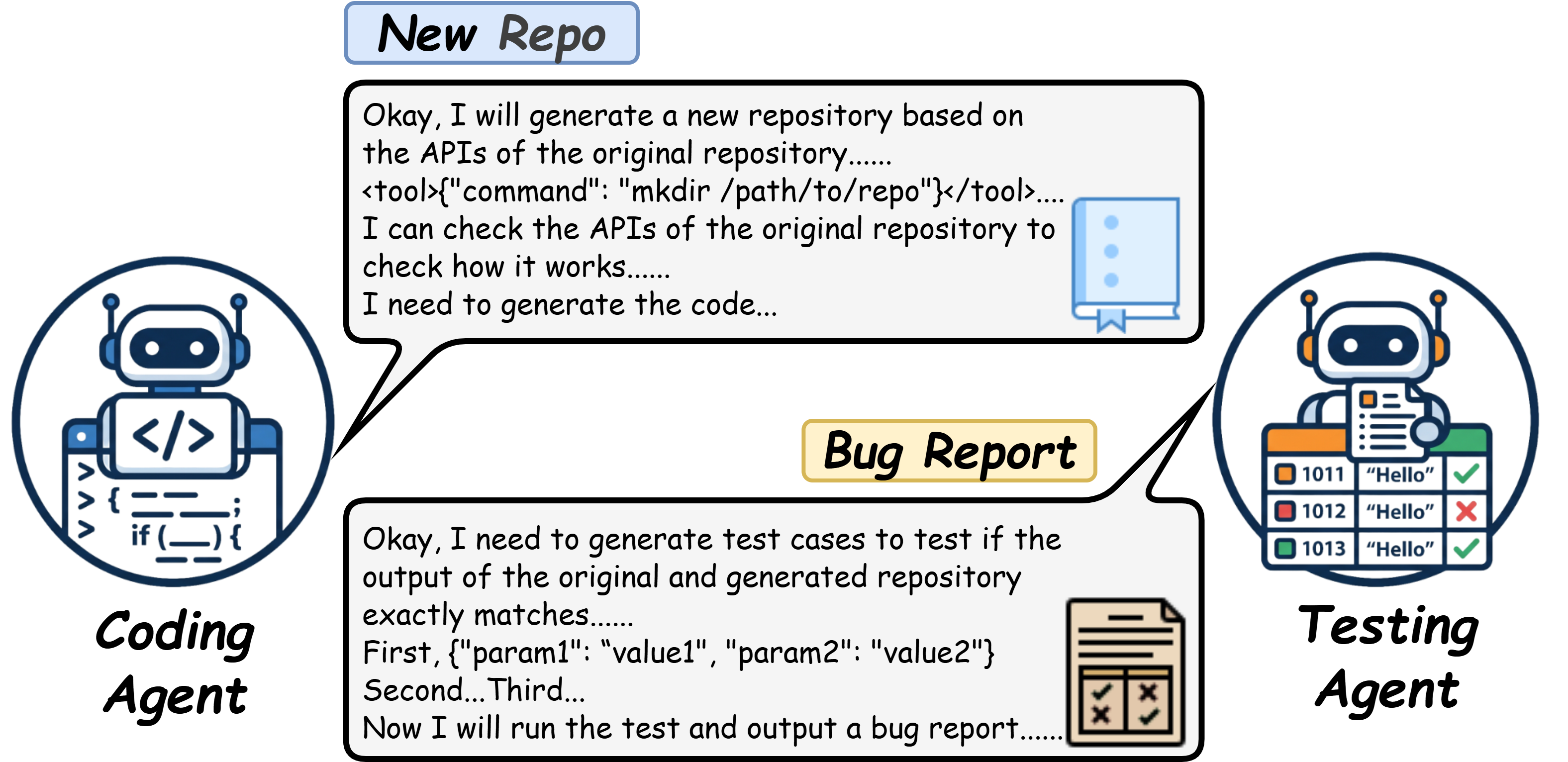}
    \end{subfigure}
    \hfill 
    \begin{subfigure}[t]{0.55\textwidth}
        \centering
        \includegraphics[width=\linewidth]{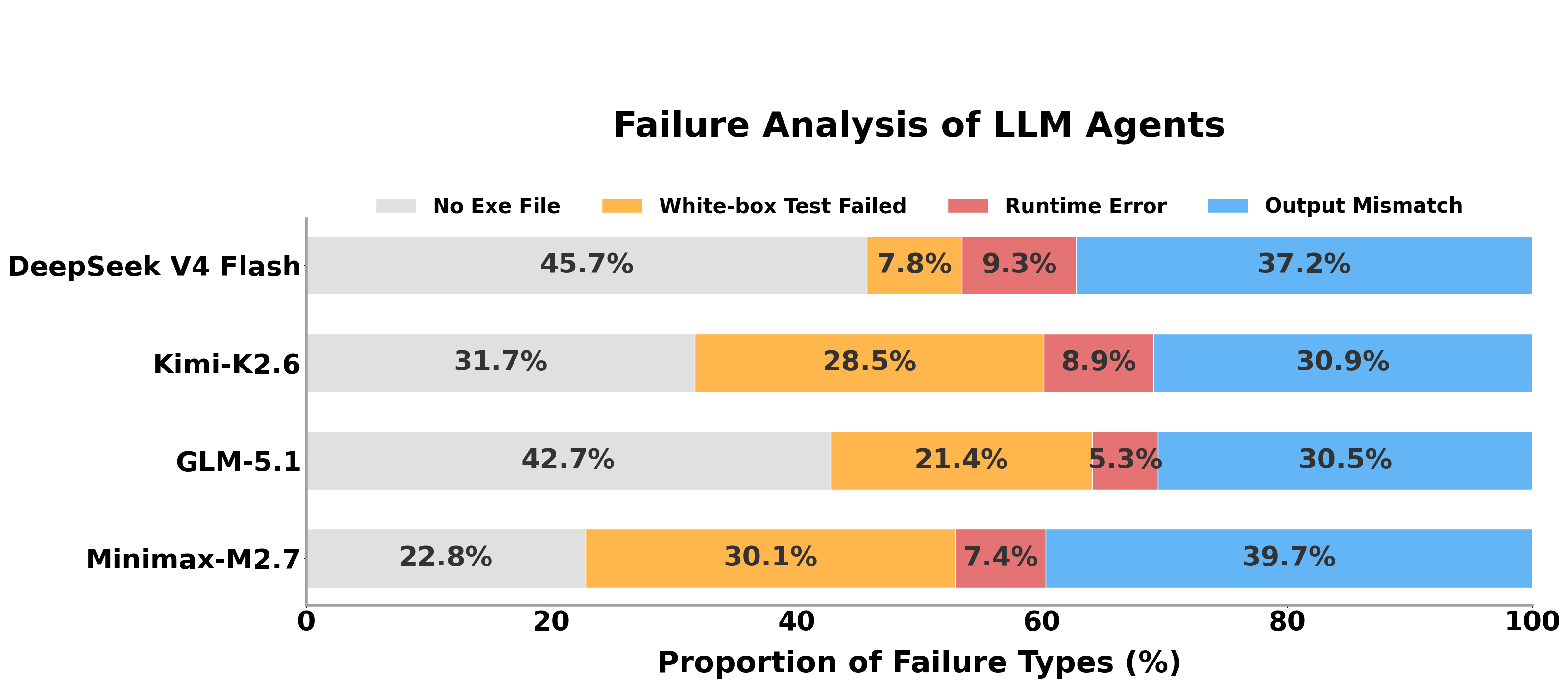}
    \end{subfigure}
    
    \caption{\textbf{Left:} an example illustrating the agentic code-test evolution framework. \textbf{Right:} analysis of failure conditions of RepoZero.}
\label{fig:loop_example}

\end{figure}

Our analysis reveals that Mini-SWE-Agent generally outperforms OpenHands-bash. This performance gap can likely be attributed to more sophisticated context engineering in Mini-SWE-Agent, whereas OpenHands-bash represents a baseline agent configuration restricted to a rudimentary bash interface. Furthermore, as illustrated in Fig.~\ref{fig:loop_example} (right), OpenHands-bash fails even on test cases explicitly provided within the initial prompt. This observation suggests that without robust, carefully designed context management, agents struggle to retain and leverage critical information during complex, long-context reasoning tasks.

\begin{figure}[b]
    \centering
    \includegraphics[width=1\linewidth]{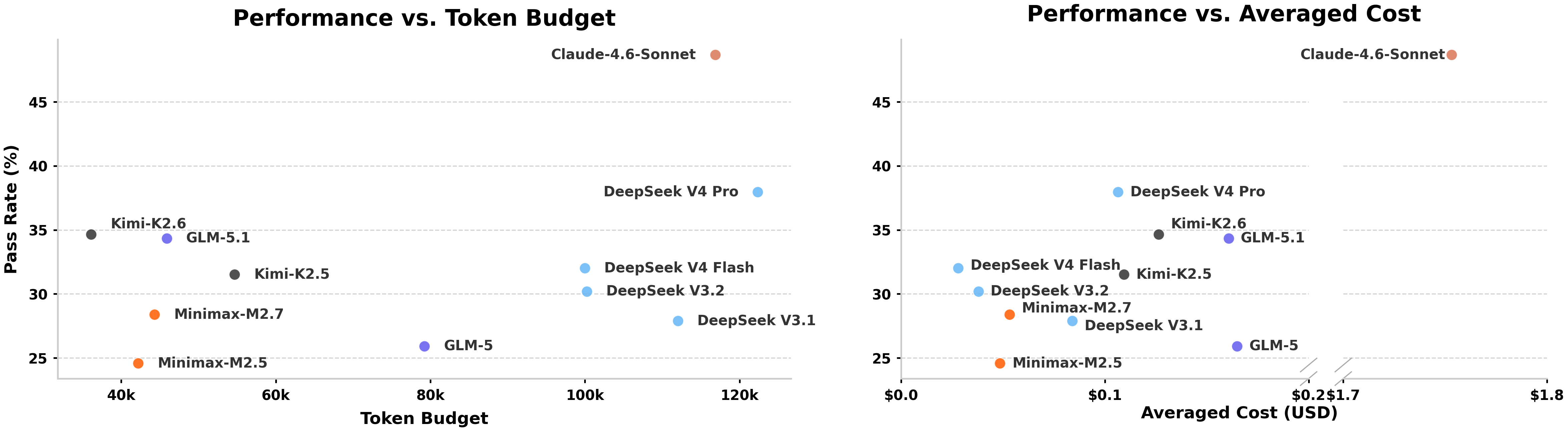}
    \caption{\textbf{Left:} pass rate vs. average token budget. \textbf{Right:} pass rate vs. average cost in USD. All models are evaluated with \includegraphics[height=0.9em]{figs/openhands.png} OpenHands-bash.}
    \label{fig:tradeoffs}
\end{figure}

\paragraph{Benchmark Validity} 
While the semi-synthetic nature of RepoZero precludes inherent quality guarantees, its empirical validity is substantiated by consistent performance alignment within established model families. Adhering to the $Benchmark^2$ principle \citep{qian2026benchmark}, our results (Tables~\ref{tab:openhands} and \ref{tab:mini-swe-agent}) demonstrate that successive iterations of the Kimi, GLM, DeepSeek, and Minimax series exhibit progressive performance gains. Furthermore, the performance hierarchy—especially Claude and Ernie—aligns with broader industry benchmarks. This structural consistency across diverse model tiers underscores the reliability of RepoZero.
\label{sec:costs}
\paragraph{Cost-Performance Trade-offs}
Figure \ref{fig:tradeoffs} illustrates the general correlation between inference expenditure and model efficacy, where the pass rate typically scales with per-sample cost. While Claude-4.6-Sonnet defines the upper bound for both metrics, this relationship is not strictly monotonic. Notably, GLM-5.1 significantly outperforms GLM-5 despite a nearly identical cost structure, highlighting how architectural advancements can decouple performance from computational overhead. However, the persistent performance gap between these leading open-source models and SOTA closed-source systems (e.g., Claude) remains a primary challenge for open-source development.

\begin{table*}[t]
\centering
\caption{Evaluation on \includegraphics[height=0.9em]{figs/icons8-python-96.png} \textbf{Py2JS} \includegraphics[height=1em]{figs/icons8-javascript-96.png} with DeepSeek V3.1 as the backbone model. We present the pass rate of models across Easy, Medium, and Hard task difficulties,  and the maximum retry times are set to $0$ (coding only), $1$ (coding-testing-refining), and $2$ (coding-testing-refining-testing-refining).}
\label{tab:tts}
\resizebox{0.85\textwidth}{!}{
\begin{tabular}{l|cccc|cccc}
\toprule
\multirow{2}{*}{\textbf{Method}} & \multicolumn{4}{c|}{\includegraphics[height=0.9em]{figs/openhands.png} \textbf{Openhands-bash}} & \multicolumn{4}{c}{\includegraphics[height=0.9em]{figs/swe-agent.png} \textbf{Mini-SWE-Agent}} \\ \cmidrule(lr){2-5} \cmidrule(lr){6-9} 
 & Easy & Medium & Hard & \textbf{Avg.} & Easy & Medium & Hard & \textbf{Avg.} \\ \midrule
\includegraphics[height=0.9em]{figs/deepseek.jpg} Retry-$0$ & 34.11 & 25.69 & 18.11 & 26.08$\pm$2.26  & 58.91& 45.14& 26.77& 43.76$\pm$2.50 \\
\includegraphics[height=0.9em]{figs/deepseek.jpg} Retry-$1$ & 45.74 & 37.50 & 25.20 & 36.15$\pm$2.38  & 58.91 & 47.92 & 29.92 & 45.67$\pm$2.44  \\
\includegraphics[height=0.9em]{figs/deepseek.jpg} Retry-$2$ & 51.14 & 41.51 & 37.11 & 42.88$\pm$2.93  & 64.34 & 51.39 & 40.16 & 52.06$\pm$2.38  \\ \bottomrule
\end{tabular}
}
\end{table*}
\subsection{Test-time Scaling via ACE}
Table~\ref{tab:tts} summarizes the ACE (a case study is provided in Fig.~\ref{fig:loop_example} left) performance of OpenHands-bash and Mini-SWE-Agent on the RepoZero-Py2JS benchmark, both of which are underpinned by the DeepSeek V3.1 backbone. The empirical results demonstrate a substantial performance gain when test cases are dynamically generated and executed during the inference phase. These findings from the ACE framework underscore two pivotal directions for the evolution of repository-level coding agents: (1) generating high-quality, executable, and verifiable test cases; and (2) integrating predefined test cases—whether human-authored or LLM-generated—in automated software engineering workflows.
\subsection{Failure Analysis}
\label{sec:fail}
We categorize the identified failure modes into four distinct groups: 
(1) failure to generate the executable code; 
(2) failure to pass white-box test cases (which are provided to the agent as illustrative examples); 
(3) general runtime errors, such as arithmetic overflow or dependency conflicts; and 
(4) output mismatches between the source and target repositories during black-box testing. 

The performance breakdown of open-source models with the OpenHands-bash scaffold is illustrated in Fig.~\ref{fig:loop_example} (right). Our analysis reveals that runtime errors occur infrequently, whereas other failure modes predominate. Based on these observations, we offer the following insights for the design of future autonomous coding agents:

\paragraph{Long-term Memory and Context Retention}
A non-negligible portion of failures was observed even within white-box test cases, a phenomenon primarily attributable to deficiencies in the agents' long-context management. Despite the inclusion of all white-box constraints in the initial prompt, agents frequently exhibit "contextual drift" during extended reasoning sequences, losing track of critical requirements as the trajectory length increases. 

\paragraph{Autonomous Environment Utilization}
We observe that most high-performing agents proactively design and execute automated unit tests within the provided environment. This self-correcting capability is instrumental in mitigating runtime errors, which are infrequent across our evaluation suite. Integrated with the empirical data in Table~\ref{tab:tts}, these findings underscore that the ability to leverage an executable environment for test-driven development is a fundamental determinant of success in complex software engineering tasks.

\paragraph{Discrepancy Between Runnability and Correctness}
While many existing benchmarks evaluate repository-level performance based solely on execution success, our results expose a substantial gap between runnability and semantic correctness. Notably, approximately 40\% of the executable code generated by agents fails to match the deterministic output of the source repository. This discrepancy reinforces the necessity of the more stringent, output-verified evaluation framework of RepoZero.
\section{Related Works}
\subsection{Repo-level Coding Benchmarks}
Traditional coding benchmarks \citep{chen2021evaluating,lu2021codexglue,gu2024cruxeval} primarily assess the capability of LLMs to generate isolated code snippets, a scope that fails to capture the complexities of real-world software engineering. To bridge this gap, SWE-bench \citep{jimenez2024swebenchlanguagemodelsresolve} introduced the first evaluation framework for repository-level bug fixing. This line of research has been further extended by Multi-SWE-bench \citep{zan2025multiswebenchmultilingualbenchmarkissue}, SWE-bench-Multimodal \citep{yang2024swe}, and SWE-bench-Pro \citep{deng2025swe}, which incorporate multiple modalities and diverse programming languages.
Complementing the focus on maintenance, FEA-bench \citep{li-etal-2025-fea} emerged as the first benchmark to evaluate LLMs' ability to implement new features within existing repositories. More recently, the challenge of from-scratch repository generation has gained significant attention. Although CodeS \citep{zan2024codes} and Commit0 \citep{zhao2024commit0} automate repository collection from GitHub, these approaches struggle with scalability and are inherently susceptible to data leakage. While NL2Repo \citep{ding2025nl2repo} and EvoCodeBench \citep{zhang2026evocodebenchhumanperformancebenchmarkselfevolving} provide manually curated natural-language descriptions and test suites for Python repositories, and E2EDevBench \citep{liu2025e2edev} utilizes GitHub-sourced repositories with "LLM-as-a-judge" rubrics, a fundamental bottleneck remains: the difficulty of constructing comprehensive test cases for repository-level generation without intensive human labor.
To address this, we present a novel benchmark that evaluates an LLM agent’s ability to generate a repository from scratch by re-implementing it based on the source APIs. Our framework represents the first benchmark that facilitates both automated test-case verification and large-scale production, ensuring both evaluation and scalability.
\subsection{LLM Coding Agents}
Following the reasoning-action loop \citep{yao2022react}, SWE-Agent \citep{yang2024sweagent}, and OpenHands \citep{wang2024openhands} are the first LLM agents to conduct coding operations within a repository. LocAgent \citep{chen2025locagent}, CoSIL \citep{jiang2025cosil}, and GraphLocator \citep{liu2025graphlocator} build a call graph for the repository and implement graph operations to assist repo-level reasoning. RepoSearcher \citep{ma2025tool}, RepoNavigator \citep{zhang2025one}, and CodeScout \citep{sutawika2026codescout} apply reinforcement learning to train the agent for bug fixing. The aforementioned agents are mainly designed for bug fixing, which is verifiable via unit tests. For code generation from scratch, as far as we know, there are only a limited number of agents that are specialized for the task. CodeS \citep{zan2024codes} and RPG \citep{luo2025rpg} perform repository generation by first designing the sketch of the repository, then filling in the code. The difficulty for rule-based verification is one of the most critical reasons that hinder the development of agents that generate repositories from scratch.
\section{Conclusion}
This paper presents RepoZero, the first scalable and verifiable benchmark for end-to-end repository generation. Unlike traditional code-completion tasks, RepoZero requires models to construct entire software projects from scratch. To ensure evaluation integrity and mitigate data contamination, we introduce a cross-language protocol across two primary subsets, \textit{C2Rust} and \textit{Py2JS}, encompassing 600 test files. 

Our experimental analysis reveals that even state-of-the-art models, supported by advanced scaffolds, achieve only moderate success (approximately 40\%), and self-verification is an important technique. These findings underscore that while the iterative "code-test" loop is a promising paradigm, a significant gap remains in agents' self-verification capabilities. Future progress in repository-level generation will necessitate a more robust integration of autonomous, high-quality test suite synthesis.

\paragraph{Limitations and Future Work} 
\label{sec:future}
We propose three primary directions to advance this field: (1) developing specialized training paradigms, such as reinforcement learning or distillation, to enhance global repository-level reasoning; (2) devising metrics and constraints to improve the architectural readability and structural integrity of generated codebases; and (3) expanding the benchmark to include large-scale, multilingual, and industry-level software development scenarios.
\bibliographystyle{plainnat}
\bibliography{NIPS/custom}
\newpage
\appendix
\section{Additional Results}
\label{sec:more_results}
Table~\ref{tab:distribution} presents the performance distribution across various repository categories, while Table~\ref{tab:testcase} provides the aggregate success rate (SR) averaged across all test cases. For clarity, we categorize the evaluation set into five functional domains: \textbf{Form} (Serialization \& Data Formats), \textbf{Crypto} (Cryptography \& Encoding), \textbf{Struc} (Data Structures \& Utilities), \textbf{Math/Sci} (Mathematics \& Science), and \textbf{Spec} (Specialized Tools).

A primary limitation of this study is the constrained evaluation of the Mini-SWE-Agent compared to the OpenHands-bash framework. Specifically, proprietary models such as GPT and Gemini were excluded from the Mini-SWE-Agent suite due to computational resource constraints and associated inference costs. We aim to expand the model coverage and provide a more comprehensive benchmark in future iterations of this work.

\paragraph{Dataset Details}
Initially, we prompt the language model to generate 60 test cases per file. After filtering, we save the successful test cases and successful test files. The remaining test files (samples that pass the filtering stage) have (about) 40  cases on average, while the minimum and maximum number of test cases per file are 20 and 60.
\begin{table*}[t]
\centering
\caption{Evaluation on \includegraphics[height=0.9em]{figs/icons8-python-96.png} \textbf{Py2JS} \includegraphics[height=1em]{figs/icons8-javascript-96.png} . We present the pass rate of models over 5 categories (see Fig.~\ref{fig:distribution} for the definition of categories).}
\label{tab:distribution}
\resizebox{0.99\textwidth}{!}{
\begin{tabular}{l|ccccc|ccccc}
\toprule
\multirow{2}{*}{\textbf{Method}} & \multicolumn{5}{c|}{\includegraphics[height=0.9em]{figs/openhands.png} \textbf{Openhands-bash}} & \multicolumn{5}{c}{\includegraphics[height=0.9em]{figs/swe-agent.png} \textbf{Mini-SWE-Agent}}  \\ \cmidrule(lr){2-6} \cmidrule(lr){7-11} 
& Form & Crypto & Struc & Math/Sci & Spec & Form & Crypto & Struc & Math/Sci & Spec  \\ \midrule
\includegraphics[height=0.9em]{figs/zhipu.jpg} GLM-5 & 58.09& 76.82& 55.96&  52.65&38.33& 49.42& 71.22& 87.67&  63.36&68.92\\
\includegraphics[height=0.7em]{figs/kimi.jpg} Kimi-K2.5 & 56.17& 59.91& 85.75&  56.14&39.94& 45.26& 58.19& 70.25&  50.25&62.16\\
\includegraphics[height=0.9em]{figs/deepseek.jpg} DeepSeek V3.1 & 35.08& 44.42& 63.88&  44.42&46.40& 54.08& 71.40& 84.96&  65.26&54.57\\
\includegraphics[height=0.9em]{figs/deepseek.jpg} DeepSeek V3.2 & 48.08& 49.11& 70.84&  42.07&55.92& 57.17& 68.92& 90.51&  61.99&67.72\\
\includegraphics[height=0.9em]{figs/ernie.png} Ernie-5.0 & 47.24& 33.06& 37.48& 42.60& 35.11& 14.87& 40.27& 48.62& 42.56& 54.66\\
\includegraphics[height=0.9em]{figs/minimax-color.png} Minimax-M2.5 & 43.30& 30.24& 35.03& 39.42& 42.36& 38.82& 57.85& 79.22& 64.88& 39.70\\
\includegraphics[height=0.9em]{figs/claude-color.png} Claude-4.6-Sonnet & 42.92& 66.26& 78.89& 67.14& 68.77& 47.10& 73.92& 88.42& 71.02& 66.79\\ \bottomrule
\end{tabular}
}
\end{table*}
\begin{table*}[t]
\centering
\caption{Evaluation on the full dataset. We present the averaged success rate across all test cases.}
\label{tab:testcase}
\resizebox{0.85\textwidth}{!}{
\begin{tabular}{l|cccc|cccc}
\toprule
\multirow{2}{*}{\textbf{Model}} & \multicolumn{4}{c|}{\includegraphics[height=0.9em]{figs/icons8-python-96.png} \textbf{Py2JS} \includegraphics[height=1em]{figs/icons8-javascript-96.png}} & \multicolumn{4}{c}{\includegraphics[height=1em]{figs/icons8-c-96.png} \textbf{C2Rust} \includegraphics[height=0.9em]{figs/icons8-rust-48.png}} \\ \cmidrule(lr){2-5} \cmidrule(lr){6-9} 
 & Easy & Medium & Hard & \textbf{Avg.} & Easy & Medium & Hard & \textbf{Avg.} \\ \midrule
\includegraphics[height=0.9em]{figs/zhipu.jpg} GLM-5 & 68.65& 51.75& 53.23& 57.67& 76.26& 65.56& 60.08& 67.27\\
\includegraphics[height=0.7em]{figs/kimi.jpg} Kimi-K2.5 & 61.21& 56.08& 59.83& 58.88& 67.36& 53.98& 49.44& 56.85\\
\includegraphics[height=0.9em]{figs/ernie.png} Ernie-5.0 & 50.76& 34.96& 32.33& 39.22& 49.07& 32.25& 34.01& 38.23\\
\includegraphics[height=0.9em]{figs/minimax-color.png} Minimax-M2.5 & 45.17& 40.04& 28.38& 37.99&  62.77&  50.80&  47.95&  53.47\\
\includegraphics[height=0.9em]{figs/deepseek.jpg} DeepSeek V3.1 & 55.93& 45.51& 36.79& 46.10& 77.15& 67.07& 52.91& 65.60\\
\includegraphics[height=0.9em]{figs/deepseek.jpg} DeepSeek V3.2 & 61.52& 54.59& 43.71& 53.57& 73.65& 70.20& 62.60& 68.90\\
\includegraphics[height=0.9em]{figs/claude-color.png} Claude-4.6-Sonnet & 67.58& 62.01& 60.12& 63.20& 73.98& 67.78& 62.32& 68.04\\ \bottomrule
\end{tabular}
}
\end{table*}
\begin{table*}[b]
\centering
\caption{Evaluation on \includegraphics[height=0.9em]{figs/icons8-python-96.png} \textbf{Py2JS} \includegraphics[height=1em]{figs/icons8-javascript-96.png} with DeepSeek V3.1 as the backbone model. We present the averaged success rate across all test cases across Easy, Medium, and Hard task difficulties, and the maximum retry times are set to $0$ (coding only), $1$ (coding-testing-refining), and $2$ (coding-testing-refining-testing-refining).}
\label{tab:ace_sr}
\resizebox{0.85\textwidth}{!}{
\begin{tabular}{l|cccc|cccc}
\toprule
\multirow{2}{*}{\textbf{Method}} & \multicolumn{4}{c|}{\includegraphics[height=0.9em]{figs/openhands.png} \textbf{Openhands-bash}} & \multicolumn{4}{c}{\includegraphics[height=0.9em]{figs/swe-agent.png} \textbf{Mini-SWE-Agent}} \\ \cmidrule(lr){2-5} \cmidrule(lr){6-9} 
 & Easy & Medium & Hard & \textbf{Avg.} & Easy & Medium & Hard & \textbf{Avg.} \\ \midrule
\includegraphics[height=0.9em]{figs/deepseek.jpg} Retry-$0$ & 55.93& 45.51& 36.79& 46.10& 77.15& 67.07& 52.19& 65..60\\
\includegraphics[height=0.9em]{figs/deepseek.jpg} Retry-$1$ & 65.85& 63.74& 54.31& 61.43& 81.21& 73.72& 61.70& 72.32\\
\includegraphics[height=0.9em]{figs/deepseek.jpg} Retry-$2$ & 67.50& 58.57& 59.69& 61.64& 78.56& 72.66& 67.53& 72.93\\ \bottomrule
\end{tabular}
}
\end{table*}
\section{A Deeper Look into ACE}
Table~\ref{tab:ace_sr} reports the success rates achieved through the ACE workflow. A comparative analysis of the \textit{Retry-1} and \textit{Retry-2} iterations reveals that while the individual test case success rate exhibits only a marginal increase, the overall sample-level pass rate improves significantly (see Table~\ref{tab:tts}). This divergence suggests that the ACE framework is particularly effective at resolving "corner case" failures within samples; by rectifying these pivotal edge cases, the framework disproportionately enhances the aggregate pass rate of entire repositories.

Building upon the empirical results of ACE, we propose two strategic directions for future research:

\begin{itemize}
    \item \textbf{Comprehensive Test Suite Synthesis}: For agent-based software development, it is imperative to construct exhaustive test suites. This enables coding agents to leverage the ACE framework for rigorous self-verification, ensuring that generated codebases adhere to complex functional requirements.
    \item \textbf{High-Fidelity Predictive Oracles}: In scenarios where large-scale test cases are difficult to procure, developing advanced "oracle models" is essential. Such models must possess the capacity to precisely predict execution outputs given specific inputs and source code. Integrating these specialized testing models into the ACE workflow could substantially augment the reasoning and self-correction capabilities of coding agents.
\end{itemize}

\clearpage
\raggedbottom
\section{Prompts}
\label{sec:prompts}
We provide the prompt templates used in our agentic evaluation and ACE loop. Placeholders enclosed by angle brackets are instantiated for each task with the corresponding source file, target path, and repository-specific dependency constraints.

\begin{promptbox}[System Prompt]
[system]
You are an AI expert with Linux terminal access. Use your reasoning ability before calling any tools.

[tool]
execute_shell(command): Execute a Linux shell command and return its output.
\end{promptbox}

\begin{promptbox}[Py2JS Generation Prompt]
[user]
You are an expert cross-language migration engineer (Python to Node.js). Read the following Python source code:

--- Source (<DATASET_ROOT>/<RELATIVE_PATH>) ---
<PYTHON_SOURCE_CODE>
---
<WHITE_BOX_TEST_CASES>
Requirements:
1. **Environment**: Write pure JavaScript for Node.js using ES Modules (ESM). Use `import`/`export`. Do NOT use `require()` or `module.exports`.
   - All generated library and entry files must use the `.mjs` suffix.

2. **CLI arguments**: The JS file must accept exactly the same command-line arguments as the Python file (same names, defaults, required fields). Parse via `process.argv`. `node test.mjs --arg val` must behave identically to `python test.py --arg val`.

3. **Logic and output**: Algorithm logic, numeric precision, and string formatting must exactly match the Python source. `console.log` output must be byte-for-byte identical to Python `print` output.

4. **Zero external dependencies**: Do NOT use any npm packages (e.g., yargs, argparse). Only import local files. Use only Node.js built-in modules (e.g., `node:fs`, `node:path`, `node:url`).

5. **Project structure**: Generate library files in `<OUTPUT_BASE>/packages/<RELATIVE_PATH_WITHOUT_PY>_pkg`. Split library code into modules with `export`. In ESM, `import` statements must include full file extensions (e.g., `import { x } from './utils.mjs'`). Working directory: `<OUTPUT_BASE>`.

6. **Black-box implementation**: Do not read Python library source. Infer behavior from the interface and re-implement in JS. Do NOT reference `<FORBIDDEN_JS_APIS_OR_LIBRARIES>`.

Output library files first (.mjs), then the entry test file `<OUTPUT_MJS_PATH>`.
\end{promptbox}

\begin{promptbox}[C2Rust Generation Prompt]
[user]
You are an expert C++ to Rust migration engineer. Read the following C++ source code:

--- Source (<CPP_SOURCE_PATH>) ---
<CPP_SOURCE_CODE>
---
<WHITE_BOX_TEST_CASES>
Requirements:
1. **Environment**: Write pure Rust using the 2021 edition. Code must compile with `rustc` or as a Cargo project.

2. **CLI arguments**: The Rust binary must accept exactly the same command-line arguments as the C++ binary (same names, defaults, and required fields). Parse args via `std::env::args()`.

3. **Logic and output**: Algorithm logic, numeric precision, and string formatting must exactly match the C++ source. `println!` output must be byte-for-byte identical to `std::cout` output.

4. **Zero external dependencies**: Do NOT use any crates from crates.io (only `std`). Forbidden crates for this repo: <FORBIDDEN_CRATES>. Implement all functionality from scratch using the standard library.

5. **Project structure**: Create a complete Cargo project in `<PACKAGE_DIR>`. Organize library code into modules. Place the entry test file `<RUST_FILENAME>` in the package root `<PACKAGE_DIR>`.

6. **Black-box implementation**: Do not read the C++ library source. Infer behavior from the interface, then re-implement using `std`.

7. **Output**:
   - Create and implement library files inside `<PACKAGE_DIR>`.
   - Save the entry file to `<OUTPUT_RUST_PATH>`.
   - Compile and produce an executable at `<OUTPUT_BINARY_PATH>`.

Hint: `<COMPILED_CPP_BINARY>` is the compiled C++ binary -- use it to debug with any arguments.
\end{promptbox}

\begin{promptbox}[ACE Test Generation Prompt]
[user]
Read the code and save test parameters to <TEST_CASES_JSON_PATH> as a JSON array:

You are a test engineer. Read the following Python code and generate 5 different sets of command-line arguments to test its logic.
Cover: normal input, boundary values, and possible error inputs.

Source code:
<PYTHON_SOURCE_CODE>

Output a JSON array of strings only, e.g.: ["--input 1", "--input 10 --verbose", ""]
No explanations, no output values -- only input argument strings.
\end{promptbox}

\begin{promptbox}[ACE Refinement Prompt Suffix]
[user]
<PY2JS_GENERATION_PROMPT>

[IMPORTANT] Do not rewrite from scratch -- fix the existing code.
Current path: <OUTPUT_MJS_PATH>
Library path: <OUTPUT_BASE>/packages/<RELATIVE_PATH_WITHOUT_PY>_pkg
Previous failure:
<FAILURE_MESSAGE>
\end{promptbox}

\newpage

\end{document}